\documentclass[11pt]{article}
\usepackage[margin=1in]{geometry}
\usepackage{amsmath,amssymb,amsthm}
\usepackage{graphicx,tikz}
\usepackage{cite}
\usepackage[hidelinks,pdfusetitle]{hyperref}

\DeclareMathOperator{\Tr}{Tr}
\DeclareMathOperator{\id}{id}
\newcommand{\ket}[1]{\lvert #1\rangle}
\newcommand{\bra}[1]{\langle #1\rvert}
\newcommand{\pr}[1]{\ket{#1}\!\bra{#1}}
\newcommand{\Rcr}{R_{\mathrm{cr}}}

\newcommand{\rmix}{r_{\mathrm{mix}}}
\newcommand{\norm}[1]{\lVert#1\rVert_1}

\newtheorem{theorem}{Theorem}
\newtheorem{lemma}[theorem]{Lemma}
\newtheorem{proposition}[theorem]{Proposition}
\newtheorem{corollary}[theorem]{Corollary}
\theoremstyle{definition}
\newtheorem{definition}[theorem]{Definition}

\title{A Koashi--Imoto Criterion for Holevo-Rate Visible Compression}
\author{Anthony M. Polloreno\thanks{E-mail: \href{mailto:ampolloreno@gmail.com}{ampolloreno@gmail.com}.}}
\date{}

\begin{document}
\maketitle

\begin{abstract}
For a finite memoryless quantum source whose encoder knows the sequence of classical labels, we determine when the Holevo rate is achievable with shared classical randomness and no shared entanglement. In the maximal Koashi--Imoto decomposition of the signal ensemble, this occurs exactly when each signal is pure on the quantum part of every sector it occupies. For qubits, this means that the signals are all pure or commute pairwise. If any occupied quantum part is mixed, every code with sufficiently small mean local error has rate bounded away from the Holevo information, uniformly in blocklength. To prove the converse, we apply Fawzi--Renner recovery to a hypothetical sequence of visible codes approaching the Holevo rate at vanishing local error, then invoke a decoupling bound of Baghali Khanian and Winter for blind compression.
\end{abstract}


\section{Introduction}
\begin{figure}[htbp]
\centering
\begin{tikzpicture}[x=1pt,y=1pt,every node/.style={font=\small,inner sep=2pt},
  box/.style={draw=black!65,rounded corners=2pt,line width=.55pt,align=center,
    minimum width=79pt,minimum height=26pt,fill=black!2},
  wire/.style={->,draw=black!80,line width=.65pt}]
  \node[box] (alice) at (95,0) {Alice\\knows $x^n$};
  \node[box] (bob) at (300,0) {Bob\\applies $\mathcal D_\nu$};
  \node[draw=black!65,rounded corners=2pt,fill=black!2,inner sep=4pt] (seed) at (197.5,27) {Shared seed $\nu$};
  \draw[wire,dashed] (seed.west) -| (alice.north);
  \draw[wire,dashed] (seed.east) -| (bob.north);
  \draw[wire] (15,0) -- (alice.west) node[midway,above] {$x^n$};
  \draw[wire] (alice.east) -- (bob.west)
    node[midway,above] {Quantum message $M$}
    node[midway,below] {$\log D=nR$ qubits};
  \draw[wire] (bob.east) -- (395,0) node[midway,above] {$W^n$};
\end{tikzpicture}
\caption{For a word $x^n$ of $n$ source labels, Alice sends a $D$-dimensional quantum register $M$ at rate $R=(\log_2D)/n$ qubits per signal. Bob applies the decoding channel $\mathcal D_\nu$ to produce $W^n$. The shared seed $\nu$ is independent of the source. For each word, we compare Bob's seed-averaged output with $\bigotimes_i\omega_{x_i}$ and average the trace distance over words.}
\label{fig:protocol}
\end{figure}
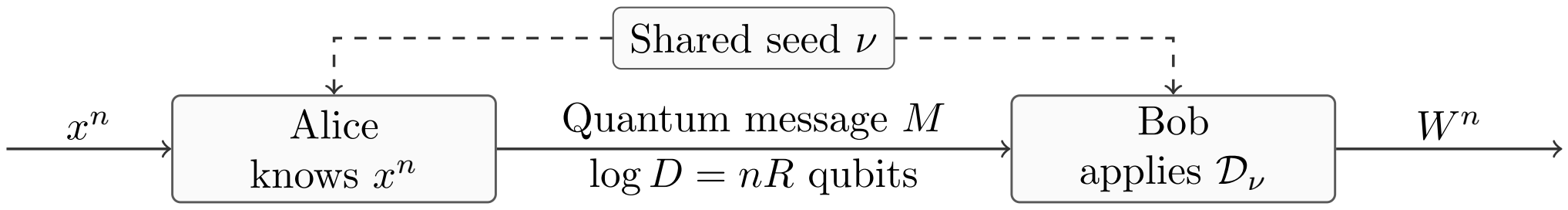

In visible compression, Alice knows the label $x$ of the state $\omega_x$ that Bob is to reproduce on the signal register $W$. We allow Alice to send a quantum message and share classical randomness with Bob, as shown in Fig.~\ref{fig:protocol}. Since a mixed state admits many pure-state decompositions, she may use the shared randomness to choose among different preparations of the same signal. Let $X$ denote the source label, drawn independently with probabilities $p_x$ on each use, and let $S$ denote von Neumann entropy in bits. The optimal rate with common randomness, denoted $\Rcr$, is measured in qubits per signal and is bounded below by the Holevo information~\cite{holevo73,horodecki,bcfjs,winter02}
\begin{equation}
 \chi=S(\bar\omega)-\sum_xp_x S(\omega_x),\qquad \bar\omega=\sum_xp_x\omega_x .
\end{equation}

For pure signals, Schumacher compression attains $\chi$~\cite{schumacher}. Commuting signals attain the same rate by classical channel simulation: Alice and Bob simulate a draw in a common eigenbasis, after which Bob prepares the indicated eigenstate~\cite{dur,winter02}. Winter~\cite[Conj.~23]{winter02} and Bennett \emph{et al.}~\cite[preprint, Conj.~27]{bennett} conjectured that shared randomness would suffice for every ensemble. One-shot lower bounds are known for compression into a subspace~\cite{jrs} and for the combined cost of communication and shared entanglement~\cite{bhn}. For blind compression, in which Alice receives $\omega_x$ without its label $x$, the Koashi--Imoto (KI) decomposition~\cite{ki} determines the optimal rate~\cite{ki01,kw}. Here we use the same decomposition to characterize when the visible rate equals $\chi$: every occupied quantum part must be pure. For every source with a mixed occupied quantum part, the converse gives a uniform gap even under the weaker local error criterion.

\begin{figure}[!b]
\centering
\includegraphics[width=.82\linewidth]{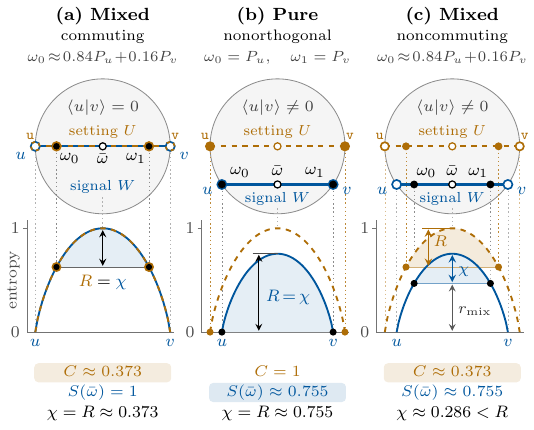}
\caption{Entropy along the chords for two equiprobable labels. In every panel $t_1=1-t_0$, with $t_0=(1-\sqrt{8/17})/2$ in (a) and (c) and $t_0=0$ in (b). The overlap $|\langle u|v\rangle|$ is zero in (a) and $2\sqrt{2}/5$ in (b) and (c). The dashed diameter represents the classical register $U$ in its own basis. The solid chord represents the physical qubit $W$. Dotted guides mark each label's entropy on the corresponding curve. The rise from the average label entropy to the peak gives $C=I(X:U)$ on the dashed curve and $\chi=I(X:W)$ on the solid curve. The highlight marks the cheaper of the two protocols shown, $R=\min\{C,S(\bar\omega)\}$. In (a) and (b), $R=\chi$. In (c), $r_{\rm mix}=S(W|X)$ and the displayed difference $R-\chi$ bounds the optimal gap $\Rcr-\chi$ from above. Theorem~\ref{thm:main} proves that the optimal gap is positive for this mixed, noncommuting pair.}
\label{fig:chords}
\end{figure}

\emph{A qubit example.} Let $P_u=\pr u$ and $P_v=\pr v$ be the pure endpoints of the off-center chord in Fig.~\ref{fig:chords}(c). Consider two distinct mixed states $\omega_x=(1-t_x)P_u+t_xP_v$, $x\in\{0,1\}$, on this chord. Direct Schumacher compression of the signals costs $S(\bar\omega)$ qubits per signal~\cite{schumacher}. Alternatively, let $U\in\{u,v\}$ record a sampled endpoint, with $\Pr(U=v\mid X=x)=t_x$. With shared randomness, Alice and Bob can simulate $U$ at a cost of $C=I(X:U)$ bits per signal, the mutual information between the source label and the endpoint~\cite{soljanin,cuff}. Alice can send each bit as one qubit in an orthogonal basis, and Bob prepares the endpoint state on $W$. Since its preparation depends only on $U$, the conditional mutual information $I(X:W|U)$ vanishes. The chain rule with $\chi=I(X:W)$ gives
\[
 C=I(X:U)=I(X:UW)=\chi+I(X:U|W).
\]
Equality $C=\chi$ would therefore imply $I(X:U|W)=0$, allowing a channel to recover the joint endpoint record and qubit from the qubit alone, for both labels~\cite{hjpw}.

Each endpoint projector is an affine combination of the two distinct mixtures. By linearity, a recovering channel would have to preserve each pure endpoint and attach its label. An isometry preserving two nonorthogonal pure states leaves the same auxiliary state for both, by preservation of their inner product~\cite[Prop.~7.2]{hjw02}. The required labels in $U$ are orthogonal, so $C>\chi$.

The endpoint protocol is the cheaper one in Fig.~\ref{fig:chords}(c), and its cost exceeds $\chi$ by about $0.087$ qubits per signal. Soljanin found a similar excess for a different qubit decomposition~\cite[Example~1]{soljanin}. The plotted excess bounds $\Rcr-\chi$ from above. Theorem~\ref{thm:main} proves $\Rcr-\chi>0$ for this pair. Since a pure decoded block may have mixed marginals, the converse retains earlier quantum outputs when examining a single position. The supplement gives the general rate formula (Theorem~\ref{thm:rate}) and an ensemble for which two copies have a strictly lower decomposition cost per signal than one (Theorem~\ref{thm:counter}).

\section{Model and equality criterion}\label{sec:model}
The ensemble defines the classical--quantum (cq) state $\sigma_{XW}=\sum_xp_x\pr{x}\otimes\omega_x$, with finite alphabet $\mathcal X$, $p_x>0$, and finite-dimensional $W$. Thus $\chi=I(X:W)_\sigma$. We use $H$ for Shannon entropy. All logarithms are base two unless stated otherwise.

For a block of $n$ independent labels, Alice sends a state $m_{x^n,\nu}$ of a $D$-dimensional register $M$ and Bob applies a decoding channel $\mathcal D_\nu$. Their shared seed $\nu$ is independent of $X^n$, and they have no other communication or shared entanglement. Averaging over the seed gives the decoded state $\tau_{X^nW^n}$ with the source prior and conditional outputs $\mathbb E_\nu\mathcal D_\nu(m_{x^n,\nu})$. With $d(a,b)=\tfrac12\lVert a-b\rVert_1$, the global and mean local errors are
\begin{equation}
 \delta=d(\tau_{X^nW^n},\sigma_{XW}^{\otimes n}),\qquad
 \delta_{\mathrm{loc}}=\frac1n\sum_i d(\tau_{X_iW_i},\sigma_{XW})\le\delta .
\end{equation}
In each one-site marginal the label $X_i$ is retained and the other labels are averaged out. Baghali Khanian, Kuroiwa, and Leung~\cite{kkl25} give a regularized rate formula for unassisted visible compression at positive mean local distortion. The code rate is $R=(\log D)/n$ qubits per signal. More precisely, $\Rcr$ is the infimum of $\limsup R$ over sequences with $n\to\infty$ and $\delta\to0$.

The entire family $\{\omega_x\}$ determines the maximal Koashi--Imoto decomposition of $W$~\cite{ki}. In the formulation with a reference system~\cite[Thm.~14]{khanian_sc}, the reference here is $X$. We restrict $W$ to the support of $\bar\omega$. Measuring the support projector at each site and replacing the complementary outcome by a fixed state on the support leaves every target invariant and contracts both errors. The same orthogonal splitting $W=\bigoplus_j Q_j\otimes N_j$ puts every signal in the form
\begin{equation}
 \omega_x=\bigoplus_j p(j|x)\,\rho_{jx}\otimes\xi_j .
 \label{eq:ki}
\end{equation}
In sector $j$, the state $\rho_{jx}$ on the quantum part $Q_j$ (the core) and the sector probability $p(j|x)$ can depend on $x$. Since the sector $J$ is readable without disturbance, Bob can discard $N_j$ and restore it locally in its fixed state $\xi_j$. The reduced source $JQ$ therefore has the same rate and Holevo information. Choose $\dim Q=\max_j\dim Q_j$ and embed each $Q_j$ isometrically into $Q$. The classical register $J$ records the sector, giving the reduced state
\[
 \sigma_{XJQ}=\sum_{x,j}p_xp(j|x)\pr{x,j}\otimes\rho_{jx}.
\]
Write $q_j=\sum_xp_xp(j|x)$. For $q_j>0$, put $p(x|j)=p_xp(j|x)/q_j$ and $\bar\rho_j=\sum_xp(x|j)\rho_{jx}$. The entropy chain rule gives
\begin{equation}
 \chi=I(X:J)+S(Q|J)-S(Q|XJ),
 \label{eq:split}
\end{equation}
where $S(Q|XJ)=\sum_{x,j}p_xp(j|x)S(\rho_{jx})$.

\begin{theorem}\label{thm:main}
With unlimited shared randomness, $\Rcr=\chi$ if and only if $\rho_{jx}$ is pure for every $(x,j)$ with $p_xp(j|x)>0$. In general,
\begin{equation}
 \chi\le \Rcr\le\chi+S(Q|XJ).
\end{equation}
If one such state is mixed, there exist $\epsilon,\delta_0>0$, depending only on the source, such that every code at every blocklength with $\delta_{\mathrm{loc}}\le\delta_0$ has $R\ge\chi+\epsilon$.
\end{theorem}

For comparison, the optimal unassisted blind rate is $R_{\rm blind}=H(J)+S(Q|J)$~\cite{ki01,kw}. Subtracting~\eqref{eq:split} gives
\[
 R_{\rm blind}-\chi=H(J|X)+S(Q|XJ).
\]
Thus either uncertainty about the sector given the label or mixedness within an occupied quantum part suffices for a blind gap. In the visible model Alice and Bob can simulate the sector at rate $I(X:J)$. Theorem~\ref{thm:main} shows that mixedness within an occupied quantum part is then necessary and sufficient for a gap.

For qubits, a noncommuting ensemble has one sector with $Q=W$, whereas a commuting ensemble has one-dimensional quantum parts. The criterion therefore explains the three examples in Fig.~\ref{fig:chords}.

Koashi and Imoto's bound~\cite{ki_incompressible} already gives equality without a shared seed when $x$ fixes $j$ and $\rho_{jx}$ is pure. Here the sector can remain random given $x$. For example, take the qutrit states $\omega_x=\tfrac12\pr e\oplus\tfrac12\pr{\psi_x}$, where $\ket e$ is an erasure flag and $\ket{\psi_0},\ket{\psi_1}$ are distinct nonorthogonal qubit states. A shared fair coin selects the erasures, and compressing the surviving pure signals costs $\chi=\tfrac12 S(\sum_xp_x\pr{\psi_x})$ per source signal. Here $H(J|X)=1$, so blind compression costs one additional qubit per signal.

\section{Proof of Theorem~\ref{thm:main}}\label{sec:proof}
\subsection{Achievability by sector simulation}
Hayden, Jozsa, and Winter combined classical channel simulation with conditional Schumacher compression for pure-state sources~\cite[Sec.~4.3]{hjw02}. We use the sector index as the simulated variable. Knowing $x^n$, Alice uses the shared seed to simulate the conditional sector law $\prod_i p(j_i|x_i)$ with Bob. They obtain the same $j^n$, and the joint law of $(X^n,J^n)$ is within total variation $\eta_n\to0$ of $p_{XJ}^{\otimes n}$. Per source signal, simulation uses $I(X:J)+o(1)$ communicated bits, sent as orthogonal states, and $H(J|X)+o(1)$ shared bits~\cite[Thm.~9 and Rem.~10]{winter02}. The bound $H(J|X)\le\sum_xp_xS(\omega_x)$ from~\eqref{eq:ki} also meets Winter's shared-randomness budget.

Under the ideal law, the positions with sector $j$ form a source with label distribution $p(x|j)$ and average state $\bar\rho_j$. Alice and Bob group the positions by sector and use Schumacher compression at any rate above $S(\bar\rho_j)$ within each group~\cite{schumacher}. For mixed $\rho_{jx}$, Alice may sample a pure state from its spectral decomposition. Averaging over that sample restores the target and can only decrease the average trace-distance error. Bob restores the original order and appends $\xi_j$ at each site.

The fraction of positions in each sector converges in probability to $q_j$. Thus, for every $a>0$, a fixed quantum register of $\lceil n[S(Q|J)+a]\rceil$ qubits suffices on typical sector strings. On the remaining strings Bob uses a fixed output. There are finitely many sectors, so the sum of their mean block errors vanishes, and substituting the simulated sector law adds at most $\eta_n$ to the global error. The total rate can therefore approach $I(X:J)+S(Q|J)=\chi+S(Q|XJ)$, which equals $\chi$ when all occupied $\rho_{jx}$ are pure.

\subsection{Converse for a general block code}
Reading the sectors of any decoded block and discarding $N_j$ can only decrease the errors. We first make the encoded messages pure by refining the shared seed~\cite[Lemma~5]{hayashi02}. For each seed, sample a spectral component for every possible source word and include this whole table in the seed. The table is independent of the actual word, and averaging reproduces the original outputs. For fixed $n$ and $D$, the decoded tuples lie in the convex hull of the compact set of tuples from pure encoding families and decoders, so Carath\'eodory's theorem gives a finite seed $\nu$.

For each sector outcome $j^n$, retain the reduced decoder's Kraus index along with $\nu$ and $j^n$ in a classical register $U$. The decoded block $Q^n$ is then pure given $X^n$ and $U$. Data processing, independence of the seed, and the $D$-dimensional Holevo bound~\cite{holevo73} give
\begin{equation}
 \begin{aligned}
 I(X^n:UQ^n)&=I(X^n:U)+S(Q^n|U)\\
 &\le I(X^n:M|\nu)\le\log D .
 \end{aligned}
 \label{eq:cost}
\end{equation}
Choose a block position $i$ uniformly, independently of the source and seed, and set $X=X_i$, $J=J_i$, and $Q=Q_i$. Thus $X$ has the original source distribution. Include the preceding quantum outputs in $Z=(i,U,Q_{<i})$, padding unused slots with a fixed pure state. By convexity the resulting state $\gamma$ satisfies $d(\gamma_{XJQ},\sigma_{XJQ})\le\delta_{\mathrm{loc}}$. We obtain
\begin{align}
 R&\ge I(X:JQZ)_\gamma,\label{eq:record}\\
 R&\ge I(X:J)_\gamma+S(Q|JZ)_\gamma.\label{eq:entropy}
\end{align}
Indeed, independence of the labels gives $I(X^n:B)\ge\sum_i I(X_i:B)$ for any $B$. With $B=UQ^n$, retaining $UQ_{<i}Q_i$ in term $i$ also retains $J_i$ within $U$. Averaging gives~\eqref{eq:record} since $X$ is independent of $i$. With $B=U$, retaining $J_i$ gives $I(X^n:U)/n\ge I(X:iJ)_\gamma\ge I(X:J)_\gamma$. Since $U$ contains $J^n$, the chain rule gives $S(Q^n|U)/n=S(Q|JZ)_\gamma$. Combining these estimates in~\eqref{eq:cost} proves~\eqref{eq:entropy}.

\subsection{A positive gap for mixed quantum parts}
We now show that a mixed quantum part prevents the code rate from approaching $\chi$ at vanishing local error. Write $\Delta=R-I(X:JQ)_\gamma$ for the excess over the decoded Holevo information. By~\eqref{eq:record}, $\Delta\ge0$. Using the mutual-information chain rule and $S(Q|JZ)=S(Q|J)-I(Q:Z|J)$, we can express~\eqref{eq:record}--\eqref{eq:entropy} as
\begin{equation}
 \begin{aligned}
 I(X:Z|JQ)_\gamma &\le \Delta,\\
 I(Q:Z|J)_\gamma &\ge S(Q|XJ)_\gamma-\Delta .
 \end{aligned}
 \label{eq:markov}
\end{equation}
As $\delta_{\mathrm{loc}}\to0$, $I(X:JQ)_\gamma\to\chi$, so $\Delta\ge0$ gives the Holevo lower bound. If a source with $S(Q|XJ)_\sigma>0$ had no uniform gap, some codes would have $\delta_{\mathrm{loc}}\to0$ and $R\le\chi+o(1)$, hence $\Delta\to0$. The first inequality would allow a channel on $JQ$ alone to approximately reproduce the joint state $XJQZ$ from $XJQ$. Since $X$ keeps track of the label and is untouched, the same channel must work for every label of positive probability. The second inequality requires this same auxiliary system $Z$ to retain correlations accounting for the mixing within a sector.

Let a superscript $j$ denote conditioning on $J=j$. In each sector of source probability $q_j>0$, the decoded probability tends to $q_j$ and $\gamma^j_{XQ}\to\sigma^j_{XQ}$. Since $I(X:Z|JQ)_\gamma$ averages nonnegative sector terms, $I(X:Z|Q)_{\gamma^j}\to0$. The sector measurement places the decoded $Q$ in the embedded space $Q_j$ given $J=j$. For each code in the sequence, the Fawzi--Renner recovery theorem~\cite{fawzi} gives a channel $\mathcal R_j:Q_j\to Q_jZ$ satisfying
\begin{equation}
 d\bigl(\gamma^j_{XQZ},(\id_X\otimes\mathcal R_j)(\sigma^j_{XQ})\bigr)\longrightarrow0 ,
 \label{eq:recover}
\end{equation}
where contractivity allows us to use the source marginal as input. Write $\beta^j_{XQZ}=(\id_X\otimes\mathcal R_j)(\sigma^j_{XQ})$. Then $\beta^j_{XQ}\to\sigma^j_{XQ}$ in trace distance.

Baghali Khanian and Winter bound the correlations left in the environment of a nearly source-preserving channel~\cite{kw}. We apply Definition~5 and Lemma~6, parts~3 and~5, of their preprint with reference system $X$. By maximality of~\eqref{eq:ki}, the conditional source $\sigma^j_{XQ}$ has a single Koashi--Imoto sector and a trivial redundant part. In this case their $J_\varepsilon(\sigma^j_{XQ})$ is the maximum of $I(Q:E')$ over isometries $Q_j\to Q_j E'$ whose $XQ$ output has fidelity at least $1-\varepsilon$ with $\sigma^j_{XQ}$. Their lemma gives $J_\varepsilon(\sigma^j_{XQ})\to J_0(\sigma^j_{XQ})=0$ as $\varepsilon\to0$.

Dilate $\mathcal R_j$ to an isometry $Q_j\to Q_j Z E$, so $ZE$ is the environment of its marginal channel to $Q_j$. A minimal environment has dimension at most $(\dim Q_j)^2$ and embeds isometrically into $ZE$, so the bound applies for every $\dim Z$. The $XQ$ marginal is $\beta^j_{XQ}$, whose fidelity with $\sigma^j_{XQ}$ tends to one. Discarding $E$ therefore gives $I(Q:Z)_{\beta^j}\to0$ by data processing. Equation~\eqref{eq:recover} transfers this limit to $\gamma^j$, since the continuity bound for $S(Q|Z)$ depends only on $\dim Q$ and the trace distance~\cite{af,winter16}. Averaging over the sectors gives $I(Q:Z|J)_\gamma\to0$, whereas~\eqref{eq:markov} requires its limit inferior to be at least $S(Q|XJ)_\sigma>0$. This proves the local gap and the global criterion.\qed

\section*{Acknowledgment}
The author thanks Chloe Rossin and Bandit Rossin Polloreno for helpful conversations. The author used ChatGPT (OpenAI) to revise the manuscript, develop and check the proof in Sec.~\ref{sec:proof}, and prepare the figures.

\clearpage
\appendix
\section*{Supplementary material}

This supplement develops the general rate formula, its structural and
entanglement interpretations, and a proof of nonadditivity.
The proof of Theorem~\ref{thm:main} in the main text is independent of these results.

\section{Model and Decomposition Cost}
In the source model of Section~\ref{sec:model}, independent labels $x\in\mathcal X$
with $p_x>0$ specify states $\omega_x$ on a $d$-dimensional space $W$.
Write $\sigma_{XW}=\sum_xp_x\pr{x}\otimes\omega_x$ for the classical--quantum (cq) state and set
$\chi=I(X:W)_\sigma$. All logarithms are binary unless stated otherwise;
$S$, $H$, and $h_2$ denote von Neumann, Shannon, and binary entropy.
The norm $\norm{\cdot}$ is the trace norm, and $D(\rho\|\sigma)$ denotes quantum relative entropy.

An $(n,D)$ code uses an independent shared seed $\nu$, message states
$m_{x^n,\nu}$ on a $D$-dimensional space, and decoding channels
$\mathcal D_\nu$ into $W^n$. Its rate is $R=(\log D)/n$ and its global
error is $\delta=\tfrac12\norm{\tau-\sigma^{\otimes n}}$, where
$\tau$ is the seed-averaged decoded cq state. Let $\Rcr$ be the infimum
of limiting upper rates along code sequences with $n\to\infty$ and
$\delta\to0$.

In the maximal Koashi--Imoto (KI) decomposition~\cite{ki} of~\eqref{eq:ki},
we call the quantum part $Q_j$ of sector $j$ its core. Write
\[
\begin{aligned}
C_Q &= S(Q|J) = \sum_j q_j S(\bar\rho_j),\\
\rmix &= S(Q|XJ) = \sum_{x,j} p_x p(j|x) S(\rho_{jx}).
\end{aligned}
\]
Here $q_j=\sum_x p_x p(j|x)$ and
$\bar\rho_j=q_j^{-1}\sum_x p_x p(j|x)\rho_{jx}$ for $q_j>0$.
Then $\chi=I(X:J)+C_Q-\rmix$. Theorem~\ref{thm:main}
gives $\chi\le \Rcr\le\chi+\rmix$, with equality $\Rcr=\chi$
exactly when $\rmix=0$.

\begin{definition}[Decomposition cost]
A decomposition consists of a finite classical variable $U$, probabilities $p(u|x)$, and states $\varphi_{xu}$ with $\omega_x=\sum_up(u|x)\varphi_{xu}$. Its cost is $I(X:U)+S(W|U)$ on
\begin{equation}
\sigma_{XUW}=\sum_{x,u}p_xp(u|x)
\pr{x}\otimes\pr{u}\otimes\varphi_{xu}.
\end{equation}
Let $F(\sigma)$ be the infimum of these costs. The same definition applies to any finite cq ensemble, including block ensembles.
\end{definition}

\begin{lemma}[Pure atoms and attainment]
\label{lem:attainment}
The minimum defining $F(\sigma)$ is attained with pure $\varphi_{xu}$ and at most $|\mathcal X|d^2+1$ values of $U$.
\end{lemma}
\begin{proof}
Write $\varphi_{xu}=\sum_kr(k|x,u)\pr{\psi_{xuk}}$. Given $u$, draw the whole tuple $V=(k_x)_{x\in\mathcal X}$ with law $\prod_xr(k_x|x,u)$, independently of $X$, and use the atom $\pr{\psi_{xu k_x}}$. This preserves the signals and gives $I(X:UV)=I(X:U)$ and $S(W|UV)\le S(W|U)$.

The pure-conditional cq states $\rho=\sum_xq_x\pr{x}\otimes\pr{\psi_x}$
form a compact set in an affine space of dimension $|\mathcal X|d^2-1$.
Put $c(\rho)=S(W)_\rho-H(X)_\rho$. A decomposition has cost
$H(X)_\sigma+\sum_up_uc(\rho_u)$, with $\sum_up_u\rho_u=\sigma$.
Thus we minimize the last coordinate of $(\sigma,t)$ in the convex
hull of $\{(\rho,c(\rho))\}$. This hull is compact and has dimension
at most $|\mathcal X|d^2$, so the minimum exists and Carath\'eodory's theorem
uses at most $|\mathcal X|d^2+1$ points.
\end{proof}

For every decomposition,
\begin{equation}
I(X:U)+S(W|U)
=\chi+I(X:U|W)+S(W|XU).
\label{eq:excess}
\end{equation}
Both excess terms are nonnegative, so $F(\sigma)\ge\chi$.
For pure atoms its cost is $I(X:UW)$: the information about the label jointly held in the flag and the signal.

\section{The Exact Global Rate}
\begin{proposition}[Codes from shared variables]
\label{prop:rec}
Every decomposition is achievable at its cost:
\begin{equation}
\Rcr\le I(X:U)+S(W|U).
\end{equation}
In particular, $\Rcr\le F(\sigma)$.
\end{proposition}

\begin{theorem}[Rate formula]
\label{thm:rate}
The sequence $F(\sigma^{\otimes n})$ is subadditive, and
\begin{equation}
\Rcr=\lim_{n\to\infty}\frac{F(\sigma^{\otimes n})}{n}
=\inf_n\frac{F(\sigma^{\otimes n})}{n}.
\end{equation}
For every $(n,D)$ code with global error $\delta\le\frac12$, putting $\varepsilon=\sqrt{2\delta}$ gives
\begin{equation}
\begin{split}
R\ge{}&\frac{F(\sigma^{\otimes n})}{n}
-2\varepsilon\log(|\mathcal X|d)\\
&-\frac{2(1+\varepsilon)}n\,
h_2\!\left(\frac{\varepsilon}{1+\varepsilon}\right).
\end{split}
\label{eq:rateconv}
\end{equation}
\end{theorem}

\subsection{Achievability}
\begin{proof}[Proof of Proposition~\ref{prop:rec}]
Use the same conditional coding scheme as in Sec.~\ref{sec:proof}~\cite{hjw02}. Simulate $U$ given $X$ at a rate approaching $I(X:U)$ bits with shared randomness~\cite[Thm.~II.1]{cuff}. Both parties know the simulated word, and its joint law with $X^n$ approaches the ideal product law in total variation. For an occupied flag $u$, the conditional source now consists of the states $\varphi_{xu}$ with probabilities $p(x|u)$. Its average state is $\bar\varphi_u=\sum_xp(x|u)\varphi_{xu}$. Grouping positions by flag and applying Schumacher compression within each group costs asymptotically $\sum_up(u)S(\bar\varphi_u)=S(W|U)$ qubits per signal. Under the ideal law, averaging over flags gives the original product signal for each word. The same typicality and trace-distance estimates as in the sector code give vanishing global error.
\end{proof}

For the KI decomposition, take $U=(j,k)$, with $k$ an eigenvector
index of $\xi_j$ sampled independently of $X$ given $j$, and atoms
$\rho_{jx}\otimes\pr{k}$ supported on sector $j$. Their cost is
$I(X:J)+C_Q=\chi+\rmix$. Thus $F(\sigma)\le\chi+\rmix$;
together with~\eqref{eq:excess}, this gives $F(\sigma)=\chi$ when
all occupied cores are pure.

\subsection{Converse and Continuity}
For the global reconstruction criterion, every code induces a
decomposition of no greater cost, and that cost is continuous in the
decoded ensemble.

\begin{lemma}[Private quantum part]
\label{lem:private}
For a state on $XUWK$ classical on $XU$, with marginal $\sigma$ on $XW$,
\begin{equation}
 F(\sigma)\le I(X:U)+S(WK|U).
\end{equation}
\end{lemma}
\begin{proof}
Apply the pure refinement of Lemma~\ref{lem:attainment} to $WK$.
It gives a refined flag $V$ with $WK$ pure given $X,V$ and
\[
 I(X:V)+S(WK|V)\le I(X:U)+S(WK|U).
\]
Measure $K$ in an orthonormal basis and call the outcome $k$.
The flag $(V,k)$ now gives a pure-atom decomposition of $\sigma$.
Its cost, written as mutual information, satisfies
\[
 F(\sigma)\le I(X:VkW)\le I(X:VKW)
 =I(X:V)+S(WK|V).
\]
The inequality is data processing under the measurement; the equality
uses purity of $WK$ given $X,V$. Combining the two displays proves
the claim.
\end{proof}

\begin{lemma}[A code induces a decomposition]
\label{lem:code}
If $\tau_{X^nW^n}$ is the seed-averaged decoded state of an $(n,D)$
code, then $F(\tau)\le\log D$.
\end{lemma}

\begin{proof}
At fixed $n,D$, the tuples of decoded states form a compact set:
they are continuous images of the compact encoding and channel parameter
sets. Carath\'eodory's theorem therefore replaces any seed average by a
finite convex combination with the same decoded states. Dilate each decoder
to $V_\nu:M\to W^nK$, padding to one environment $K$.
Lemma~\ref{lem:private}, with $X=X^n$ and $U=\nu$, gives
$F(\tau)\le I(X^n:\nu)+S(W^nK|\nu)\le\log D$.
Indeed, the seed is independent of $X^n$, and the state of $W^nK$
given $\nu$ is an isometric image of the average message.
\end{proof}

\begin{lemma}[Continuity at a fixed prior]
\label{lem:cont}
Let $\sigma$ and $\sigma'$ be ensembles on $W$ with the same labels
and probabilities, and let
$\delta=\tfrac12\norm{\sigma-\sigma'}\le\tfrac12$.
With $\varepsilon=\sqrt{2\delta}$,
\begin{equation}
 \begin{split}
 |F(\sigma)-F(\sigma')|\le{}&2\varepsilon\log(|\mathcal X|\dim W)\\
 &+2(1+\varepsilon)h_2\!\left(\frac{\varepsilon}{1+\varepsilon}\right).
 \end{split}
 \label{eq:cont}
\end{equation}
\end{lemma}

\begin{proof}
For any decomposition of $\sigma'$, write
$\tau'_x=\sum_up'(u|x)\pr u\otimes\varphi'_{xu}$ and choose
a purification $\ket{\Psi'_x}$ on $UWE$, taking $E$ large enough.
This also purifies $\omega'_x$. Put
$\delta_x=\tfrac12\norm{\omega_x-\omega'_x}$, so
$\sum_xp_x\delta_x=\delta$. Uhlmann's theorem and the
fidelity--trace-distance inequalities~\cite{fvdg} give a purification
$\ket{\Psi_x}$ of $\omega_x$ on the same space satisfying
\[
 \tfrac12\norm{\pr{\Psi_x}-\pr{\Psi'_x}}
 \le\sqrt{2\delta_x}.
\]
Trace out $E$ and dephase $U$ in both purifications. Applied to $\ket{\Psi'_x}$,
these operations return $\tau'_x$, which is classical on $U$. Applied to
$\ket{\Psi_x}$, they give a cq extension $\tau_x$ of $\omega_x$.
By contractivity, the two extensions on $XUW$ have trace distance at most
$\sum_xp_x\sqrt{2\delta_x}\le\varepsilon$.

Their costs are $H(X)-H(X|U)+S(W|U)$, with identical $H(X)$.
Applying the conditional-entropy continuity bound~\cite{winter16}
to $X$ and $W$ gives the right side of~\eqref{eq:cont}; neither bound
depends on $\dim U$. Minimize over the decomposition of $\sigma'$
and interchange the two ensembles.
\end{proof}

\begin{proof}[Proof of Theorem~\ref{thm:rate}]
Product decompositions give
$F(\sigma^{\otimes(n+m)})\le F(\sigma^{\otimes n})+F(\sigma^{\otimes m})$.
The nonnegative subadditive sequence therefore satisfies
\[
 \lim_{n\to\infty}\frac{F(\sigma^{\otimes n})}{n}
 =\inf_n\frac{F(\sigma^{\otimes n})}{n}.
\]
Proposition~\ref{prop:rec}, applied to blocks of $n$ letters, bounds
$\Rcr$ above by every term of this infimum.

Conversely, an $(n,D)$ code with global error $\delta\le1/2$ has
$\tfrac12\norm{\tau-\sigma^{\otimes n}}=\delta$ and
$F(\tau)\le nR$ by Lemma~\ref{lem:code}. Apply Lemma~\ref{lem:cont}
with alphabet $\mathcal X^n$ and signal space $W^n$, then divide by $n$.
This is the finite-error bound~\eqref{eq:rateconv}. Its correction
vanishes with $\delta$, uniformly for $n\ge1$, proving the reverse
inequality and the rate formula.
\end{proof}

\section{Structure of the Decomposition Cost}

Two consequences of Lemma~\ref{lem:private} locate the possible savings
from joint coding in the mixed quantum cores.

\begin{proposition}[Partial additivity]
For any two ensembles,
\begin{equation}
 \chi(\sigma_1)+F(\sigma_2)
 \le F(\sigma_1\otimes\sigma_2)
 \le F(\sigma_1)+F(\sigma_2).
\end{equation}
In particular, $F$ is additive on a product if either factor has only
pure occupied cores.
\end{proposition}
\begin{proof}
For any decomposition $U$ of the product source, independence of the
labels and the chain rule give
\begin{align*}
 I(X_1X_2:U)+S(W_1W_2|U)
 &=I(X_2:UX_1)+S(W_1W_2|UX_1)+I(X_1:UW_1W_2)\\
 &\ge F(\sigma_2)+\chi(\sigma_1).
\end{align*}
The first two terms bound $F(\sigma_2)$ by Lemma~\ref{lem:private},
using $UX_1$ as the flag and $W_1$ as the private system. The last
term is at least $I(X_1:W_1)=\chi(\sigma_1)$ by data processing.
Minimizing proves the lower bound; product decompositions give the upper bound.
If either factor has only pure occupied cores, its KI construction has
cost $\chi$, so the bounds coincide, interchanging the factors if needed.
\end{proof}

\begin{proposition}[Sector splitting]
For the decomposition~\eqref{eq:ki}, let
$\sigma_j=\{p(x|j),\rho_{jx}\}$ be the core ensemble, with
$p(x|j)=p_xp(j|x)/q_j$. Then
\begin{align}
 F(\sigma)&=I(X:J)+\sum_jq_jF(\sigma_j),\\
 F(\sigma)-\chi(\sigma)
 &=\sum_jq_j\bigl[F(\sigma_j)-\chi(\sigma_j)\bigr].\nonumber
\end{align}
Maximality of the decomposition is not needed for these identities.
\end{proposition}
\begin{proof}
For the upper bound, combine decompositions of the core ensembles with
the sector $j$ and an eigenvector index of $\xi_j$, sampled independently
of $x$ given $j$. For the lower bound, start with pure atoms and measure the sector $j$,
recording the outcome in the flag. Each conditional atom remains pure,
and the source average is unchanged. The cost cannot increase: it changes
from $I(X:UW)$ to $I(X:UJW)$ under a channel on $W$.
Writing the new cost explicitly gives
\[
 I(X:J)+\sum_jq_j\bigl[I(X:U|J=j)+S(Q_jN_j|U,J=j)\bigr]
 \ge I(X:J)+\sum_jq_jF(\sigma_j).
\]
Here Lemma~\ref{lem:private} is applied within each sector with
private system $N_j$. Finally,
$\chi(\sigma)=I(X:J)+\sum_jq_j\chi(\sigma_j)$ gives the second identity.
\end{proof}

Thus all excess over the Holevo quantity lies in the cores.

\section{An Entanglement Distillation Interpretation}

Let $\sigma_{XE}$ be complementary to $\sigma_{XW}$ in any purification
$\ket{\Psi}_{XWE}$.
Although $X$ is classical in
$\sigma_{XW}$, it need not be classical in this complementary state.
Write $D_{A\to B}$ for one-way distillable entanglement with classical
communication from $A$ to $B$, and $D_\leftrightarrow$ for the two-way
quantity, all in ebits.

A purification may store a copy of the classical label in $E$ along with a purification of each signal. Up to a local isometry on $E$, this puts $\sigma_{XE}$ in premeasurement form, to which Coles's distillation identity applies~\cite[Thm.~4]{coles}. For communication from $E$ to $X$, we specialize the one-way formula of Devetak and Winter~\cite{dw} to the same state.

\begin{corollary}
For the above complementary state,
\begin{align}
 \Rcr&=H(X)-D_{E\to X}(\sigma_{XE}),\label{eq:distill}\\
 D_{X\to E}(\sigma_{XE})
 &=D_\leftrightarrow(\sigma_{XE})=H(X)-\chi,\\
 \Rcr-\chi
 &=D_{X\to E}(\sigma_{XE})-D_{E\to X}(\sigma_{XE}).
\end{align}
\end{corollary}
\begin{proof}
\emph{Communication from $E$ to $X$.}
Let $D^{(1)}_{E\to X}$ be the one-copy instrument optimization in the
Devetak--Winter formula~\cite[Thm.~13]{dw}. Refining an instrument on $E$
to one Kraus operator per outcome can only increase its average coherent
information, by concavity of conditional entropy. The conditional global states are then pure, and the
objective is
\[
 \sum_up_u\bigl[S(X)_{\rho^u}-S(W)_{\rho^u}\bigr],
 \qquad \sum_up_u\rho^u_{XW}=\sigma_{XW}.
\]
Conversely, choose a purification $\ket{\psi_u}_{XWE'}$ of each
$\rho^u$. The vector $\sum_u\sqrt{p_u}\ket u\ket{\psi_u}$
purifies $\sigma_{XW}$. Taking $E'$ large enough, uniqueness of purification
gives an isometry $E\to UE'$. Measuring $U$ then realizes the decomposition.
Dephasing $X$ in each
$\rho^u$ preserves their average and their $W$ marginals, while increasing
$S(X)_{\rho^u}$. Realize these new components by a new instrument on $E$.
Hence the optimization may be restricted to cq
decompositions, and
$D^{(1)}_{E\to X}(\sigma_{XE})=H(X)-F(\sigma)$.
Regularizing this equality and applying Theorem~\ref{thm:rate} proves
\eqref{eq:distill}.

\emph{Communication from $X$ to $E$.}
The hashing inequality~\cite{dw} achieves the coherent information $S(E)-S(XE)=H(X)-\chi$ in this direction. For a premeasurement state, Coles's theorem~\cite[Thm.~4]{coles} gives the same value for the relative entropy of entanglement, which bounds two-way distillation from above~\cite{rains}. This proves the second equality, and subtraction gives the third.
\end{proof}

Theorem~\ref{thm:main} therefore characterizes when these two directions
have equal distillable entanglement. The one-copy formula $F$ is optimal exactly
when $D_{E\to X}(\sigma_{XE})=D^{(1)}_{E\to X}(\sigma_{XE})$.

\section{Regularization Is Necessary}

The following implication will turn an entropy witness into strict
nonadditivity for one ensemble. Its proof, including label-dependent
supports, is in Appendix~\ref{app:product}.

\begin{lemma}[Product optimality]
\label{lem:product}
If $F(\sigma_1\otimes\sigma_2)=F(\sigma_1)+F(\sigma_2)$, then every
state $\tau$ classical on $X_1X_2$, with marginals $\sigma_1,\sigma_2$
and pure conditional signals, satisfies
\begin{equation}
 F(\sigma_1)+F(\sigma_2)
 \le I(X_1:X_2)_\tau+S(W_1W_2)_\tau.
 \label{eq:product-witness}
\end{equation}
\end{lemma}

We adapt random unitary methods for nonadditivity of minimum output
entropy~\cite{hastings,collins16}.
For visible compression without shared randomness or entanglement,
Horodecki~\cite{horodecki00} expressed the optimal rate as a regularized minimization over extensions of the signals. Hayashi~\cite{hayashi06} identified this rate with
$\lim_{n\to\infty}E_P(\sigma_{XW}^{\otimes n})/n$, where $E_P$ is the
entanglement of purification. Chen and
Winter~\cite{chen_winter_ep} gave numerical evidence of nonadditivity of
$E_P$, and Baghali Khanian and Negari~\cite{khanian_negari_ep} proved it
for a range of separable Werner states. Related collective advantages occur
in static communication--entanglement trade-offs even when their optimized
coherent-information value remains additive~\cite{polloreno_selective}.

\begin{theorem}[Nonadditivity]
\label{thm:counter}
For all sufficiently large integers $k$, there are a finite dimension $D$
and real orthogonal matrices $O_1,\ldots,O_k$ on $\mathbb C^D$ such that the following
ensemble satisfies $F(\sigma^{\otimes2})<2F(\sigma)$.
Set $p=24/\ln k$ and take uniform labels $x=(i,s)\in\{1,\ldots,k\}\times\{\pm1\}$.
For $v\in\mathbb C^D$, the isometry $V_x:\mathbb C^D\to\mathbb C^2\otimes\mathbb C^D$
and the signal $\omega_x$ are given by
\[
 V_xv=\sqrt{1-p}\ket{0}\otimes v+\sqrt p\ket{1}\otimes sO_iv,
 \qquad \omega_x=V_xV_x^\dagger/D.
\]
Consequently $\Rcr(\sigma)<F(\sigma)$, so the regularization in
Theorem~\ref{thm:rate} cannot in general be removed.
\end{theorem}

\begin{proof}
We use natural logarithms throughout this proof; all entropies, relative
entropies, and $F$ are rescaled together. Write
$h(t)=-t\log t-(1-t)\log(1-t)$.

\emph{The channel.}
Collins's theorem~\cite[preprint, Thm.~3.1]{collins16} supplies, for each fixed $k$
in a sufficiently large finite dimension, unitaries $U_i$ for which
$N_U(\rho)=k^{-1}\sum_iU_i\rho U_i^\dagger$ satisfies
\begin{equation}
 \Tr N_U(\rho)^2\le\frac1k\left(1+\frac{10}{k}\right)
 \quad\text{for every state }\rho.
 \label{eq:purity}
\end{equation}
Indeed, its complementary Gram matrix has Hilbert--Schmidt distance at
most $(3+o(1))/k$ from $I/k$, uniformly over pure inputs as the dimension
tends to infinity. It has the same nonzero spectrum as $N_U(\rho)$ for
pure inputs, and purity is convex. Let $O_i$ be the realification of $U_i$
in twice the dimension, denoted by $D$. Under the equivalence to
$U_i\oplus\bar U_i$, its complementary Gram
matrix is a convex combination of the corresponding Gram matrices for
$U$ and $\bar U$. Thus $N(\rho)=k^{-1}\sum_iO_i\rho O_i^\dagger$
satisfies the same purity bound, still for every complex input.

\emph{One copy.}
Every pure atom for label $x$ lies in $\operatorname{ran}V_x$. Fix one
conditional ensemble of a pure-atom decomposition, with label weights
$q_x$, unit vectors $\psi_x$, and average
$\rho=\sum_xq_xV_x\pr{\psi_x}V_x^\dagger$. Put
\[
 \rho'=\sum_xq_x\pr{\psi_x},\qquad
 \tau=\sum_{i,s}q_{i,s}O_i\pr{\psi_{i,s}}O_i^\dagger,
\]
and let $D_q=D(q\|u)$, where $u$ is uniform on the $2k$ labels.
The conditional contribution to the cost is $D_q+S(\rho)$.
Abbreviate $A=D_q+S(\rho')$.
Compare the cq states with blocks $q_x\pr{\psi_x}$ and $u_x\rho'$.
Their relative entropy is $A$. Applying the controlled
unitary $O_i$ and forgetting the label gives
$D(\tau\|N(\rho'))\le A$ by data processing. With $\eta=N(\rho')$, the identity
\[
 S(\tau)+2D(\tau\|\eta)
 =D\!\left(\tau\middle\|\frac{\eta^2}{\Tr\eta^2}\right)
 -\log\Tr\eta^2
\]
and~\eqref{eq:purity} therefore imply
\begin{equation}
 S(\tau)+2A\ge\log k-\log(1+10/k).
 \label{eq:purity-entropy}
\end{equation}

Let $\Delta$ dephase the first, two-dimensional factor and set
$C(\rho)=D(\rho\|\Delta\rho)$. Then
\[
 S(\rho)=h(p)+(1-p)S(\rho')+pS(\tau)-C(\rho).
\]
Group labels into sign pairs $(i,+),(i,-)$ with weights $w_i=q_{i,+}+q_{i,-}$.
Omit pairs with $w_i=0$. In each remaining pair set $\theta_i=q_{i,+}/w_i$
and $c_i=|\langle\psi_{i,+}|\psi_{i,-}\rangle|$, choosing an arbitrary unit
vector for any zero-weight label. Let $s(\theta,c)$ denote the entropy of a
mixture of two pure states with weights $\theta,1-\theta$ and absolute
overlap $c$. The pair has coherence
$h(p)+s(\theta_i,c_i)-s(\theta_i,(1-2p)c_i)$.
Moreover, the label chain rule and entropy concavity give
\[
 A\ge
 \sum_iw_i[\log2-h(\theta_i)+s(\theta_i,c_i)].
\]
Joint convexity of relative entropy and Lemma~\ref{lem:scalar} now give
\begin{equation}
 C(\rho)\le(1-3p)A+\varepsilon(p),
 \label{eq:flag}
\end{equation}
for $0<p\le1/64$, where $\varepsilon(p)=3p(4p)^{1/(4p)}$.
Substituting~\eqref{eq:flag} gives
\[
 D_q+S(\rho)\ge h(p)+p[S(\tau)+2A]
 +pD_q-\varepsilon(p).
\]
Apply~\eqref{eq:purity-entropy}, discard $pD_q\ge0$, and average to obtain
\begin{equation}
 F(\sigma)\ge h(p)+p\log k-p\log(1+10/k)-\varepsilon(p).
 \label{eq:counter-one}
\end{equation}

\emph{A two-copy witness.}
We use the maximally entangled input of the conjugate-channel construction~\cite{hw08,hastings}. In the basis where the $O_i$ are real, put
$\ket{\Omega}=D^{-1/2}\sum_{r=1}^D\ket{r,r}$.
Let $\tau$ be the cq state with independent uniform labels $x,y$ and
conditional pure signals $(V_x\otimes V_y)\ket{\Omega}$.
Both marginal ensembles are $\sigma$.
The independent signs dephase both flags, giving four blocks of weights
$(1-p)^2,p(1-p),p(1-p),p^2$. The first is pure; the middle blocks are
mixtures of $k$ pure states. In the last block, the $k$ equal-index
terms coincide because $(O_i\otimes O_i)\ket{\Omega}=\ket{\Omega}$.
After grouping them, one outcome has probability $1/k$ and the other
$k^2-k$ outcomes each have probability $1/k^2$. Their Shannon entropy
$(2-1/k)\log k$ bounds this block's entropy. Hence
\begin{equation}
 S(W_1W_2)_\tau\le2h(p)+2p\log k-p^2(\log k)/k.
 \label{eq:counter-two}
\end{equation}

\emph{Comparison.}
For $p=24/\log k$, eventually $p\le1/64$, and
\[
 \frac{\varepsilon(p)}{p/k}
 =3k(4p)^{1/(4p)}
 =3\exp\!\left[\left(1-\frac{\log(\log k/96)}{96}\right)\log k\right]\longrightarrow0.
\]
Using $\log(1+10/k)\le10/k$ in
\eqref{eq:counter-one}--\eqref{eq:counter-two} gives
\[
 2F(\sigma)-S(W_1W_2)_\tau
 \ge\frac{4p}{k}-2\varepsilon(p)>0
\]
for sufficiently large $k$. Since the witness labels are independent,
Lemma~\ref{lem:product} rules out product equality, and subadditivity
gives $F(\sigma^{\otimes2})<2F(\sigma)$. After restoring binary units,
Theorem~\ref{thm:rate} gives the rate inequality.
\end{proof}

By Lemma~\ref{lem:cont}, the strict inequality persists under a
sufficiently small perturbation
$\omega_x\mapsto(1-\delta)\omega_x+\delta I/(2D)$. Thus the signals
can also be chosen real and full rank, with a common spectrum.
The construction uses large dimension. Additivity for the qubit pair in Fig.~\ref{fig:chords}(c), whose Bloch vectors can be taken as $0.8\hat z$ and $0.8\hat x$, would require a separate argument.

\section{A Scalar Coherence Estimate}

All logarithms in this section are natural. Let
$h(t)=-t\log t-(1-t)\log(1-t)$, with the endpoint convention
$0\log0=0$, and let $s(\theta,c)$ be the von Neumann entropy of a mixture of two pure states
with weights $\theta,1-\theta$ and absolute overlap $c$.

\begin{lemma}
\label{lem:scalar}
For $0<p\le1/64$, set $\varepsilon(p)=3p(4p)^{1/(4p)}$.
For every $\theta,c\in[0,1]$,
\begin{equation}
\begin{split}
 h(p)+s(\theta,c)-s(\theta,(1-2p)c)
 \le{}&(1-3p)\bigl[\log2-h(\theta)+s(\theta,c)\bigr]\\
 &+\varepsilon(p).
\end{split}
 \label{eq:scalar}
\end{equation}
\end{lemma}
\begin{proof}
Put $a=1-2p$ and $b=1-a^2=4p(1-p)$. Let $\lambda$ be the smaller eigenvalue of the mixture on the span of the two states, taking $\lambda=0$ if they coincide, and put $z=(2\theta-1)^2$.
Then $0\le\lambda\le(1-\sqrt z)/2$ and
\[
 s(\theta,c)=h(\lambda),\qquad
 s(\theta,ac)=d(\lambda,z):=
 h\!\left(\frac{1-\sqrt{a^2(1-2\lambda)^2+bz}}2\right).
\]

\emph{Label imbalance.}
Set
\[
 A(z)=h(p)-d(0,z),\qquad
 B(z)=\log2-h\!\left(\frac{1-\sqrt z}2\right)=\log2-h(\theta).
\]
For $g(z)=\operatorname{artanh}(\sqrt z)/\sqrt z$, with $g(0)=1$,
\[
 A'(z)=\tfrac b2g(a^2+bz),\qquad B'(z)=\tfrac12g(z),\qquad
 g'(z)=\frac{(1-z)^{-1}-g(z)}{2z}\le\frac1{2(1-z)}.
\]
The last inequality uses $g(z)\ge1$. Integrating from $z$ to
$a^2+bz$ gives
\[
 g(a^2+bz)\le g(z)+\tfrac12\log(1/b)
 \le[1+\tfrac12\log(1/b)]g(z).
\]
Since $A(0)=B(0)=0$, we have $A\le\alpha B$, where
$\alpha=b[1+\tfrac12\log(1/b)]$. The function defining $\alpha$
is increasing for $0<b\le1$, and $b\le1/16$, so
\[
 \alpha+3p\le\frac{1+2\log2}{16}+\frac3{64}<1.
\]
Thus $A(z)\le(1-3p)B(z)$.

\emph{Mixedness.}
For admissible $0\le\lambda\le p$, put
$r=\sqrt{a^2(1-2\lambda)^2+bz}$. Since $1-2\lambda\ge a$,
$a^2\le r\le1$, and $a^3\ge3/4$,
\[
 \partial_\lambda d
 =\frac{a^2(1-2\lambda)}r\log\frac{1+r}{1-r}
 \ge a^3\log\frac{1+a^2}{1-a^2}
 \ge L_p:=\tfrac34\log\frac1{4p}.
\]
Monotonicity of $d$ in $\lambda$ now gives
$d(\lambda,z)-d(0,z)\ge L_p\min\{\lambda,p\}$ throughout the
admissible interval. Subtracting the first term on the right of
\eqref{eq:scalar} from its left side leaves at most
\[
 3ph(\lambda)-L_p\min\{\lambda,p\}.
\]
For $\lambda\le p$, this is at most
\[
 \max_{0\le t\le1}\{3ph(t)-L_pt\}
 =3p\log(1+e^{-L_p/(3p)})
 \le3p(4p)^{1/(4p)}.
\]
For $\lambda\ge p$, it is at most $3p\log2-L_pp\le0$, since
$L_p\ge3\log2$. Endpoints follow by continuity.
\end{proof}

\section{Proof of the Product-Optimality Lemma}
\label{app:product}

Shor~\cite[Sec.~6]{shor} and Pomeransky~\cite[Sec.~IV]{pomeransky} used supporting functionals to deduce strong superadditivity from additivity for entanglement of formation. We use their argument for the decomposition cost $F$. At optimal product atoms the first derivatives of a supporting bound determine it in terms of the two marginal bounds.
The argument works in either logarithm base, used consistently.
Omit zero-prior labels and put
$S_x=\operatorname{supp}\omega_x$, $T_y=\operatorname{supp}\omega'_y$ for the two ensembles.
We prove Lemma~\ref{lem:product} on these label-dependent supports.

Write an optimal pure-atom decomposition of the first ensemble as
$z_{xu}=\sqrt{p(x,u)}\ket{\varphi_{xu}}\in S_x$, and let $Z_u$ have
these vectors as columns. For blocks $A_x=p_x\omega_x$, define
\begin{align}
 G(A)&=\min_{\sum_uz_{xu}z_{xu}^\dagger=A_x}\sum_ug(Z_u),
 \\
 g(Z)&=-\Tr ZZ^\dagger\log ZZ^\dagger
       +\sum_x\|z_x\|^2\log\|z_x\|^2.\nonumber
\end{align}
For normalized blocks, $G(A)=F(\sigma_1)-H(X_1)$. Attainment follows
from Lemma~\ref{lem:attainment}. The same formula defines
a finite convex, positively homogeneous function on the cone of positive
blocks on the fixed spaces $S_x$. Indeed, with $m$ labels and
$T=\sum_x\Tr A_x$,
\[
 -T\log m\le G(A)\le0;
\]
concatenating and scaling decompositions proves convexity and
homogeneity.

Each $A_x$ is positive definite on $S_x$, so $A$ lies in the relative
interior of this cone and admits a supporting functional. Homogeneity
then gives Hermitian operators $K_x$ on $S_x$ such that
\begin{align}
 G(A)&=\sum_x\Tr K_xA_x,\nonumber\\
 g(Z)&\ge\sum_xz_x^\dagger K_xz_x
 \quad(z_x\in S_x).
 \label{eq:support-functional}
\end{align}
To see the normalization, apply the supporting inequality at $tA$ for
$t$ on either side of one. The second inequality follows by testing
$G$ at the rank-one blocks $(z_xz_x^\dagger)_x$.

Every optimal atom saturates~\eqref{eq:support-functional}: its nonnegative
slack has average zero. The function
$g$ is continuously differentiable even at singular $ZZ^\dagger$, since
$s^2\log s^2$ is continuously differentiable at zero. The difference between the atom cost and its linear bound has a minimum
of zero there. Differentiating it under variations in $S_x$ gives
\begin{equation}
 K_xz_x=P_x\bigl[(\log\|z_x\|^2-\log ZZ^\dagger)z_x\bigr],
 \label{eq:projected-stationarity}
\end{equation}
where $P_x$ projects onto $S_x$. Matrix logarithms act on their supports;
at zero columns the gradient terms have their continuous value zero.
Choose a supporting family $K'_y$ and optimal matrices $Z'_v$ for the
second ensemble in the same way.

If $F$ is additive on the product, the product matrices
$Z_u\otimes Z'_v$ are optimal for it. A supporting family $J_{xy}$
exists on $S_x\otimes T_y$. Applying~\eqref{eq:projected-stationarity}
to the product atoms, and using the tensor-product logarithm on the
support, gives
\[
 J_{xy}(z_{xu}\otimes z'_{yv})
 =(K_x\otimes I_{T_y}+I_{S_x}\otimes K'_y)(z_{xu}\otimes z'_{yv}).
\]
Because $\sum_uz_{xu}z_{xu}^\dagger=A_x$ is positive definite on
$S_x$, the vectors $z_{xu}$ span $S_x$. Likewise the $z'_{yv}$ span $T_y$.
Consequently $J_{xy}=K_x\otimes I_{T_y}+I_{S_x}\otimes K'_y$ on
$S_x\otimes T_y$.

Write the witness's conditional vectors as
$\sqrt{q_{xy}}\ket{\psi_{xy}}$. Positivity of the marginal
decompositions forces them to lie in $S_x\otimes T_y$.
Applying~\eqref{eq:support-functional} for $J$ to this one atom and
using its prescribed marginals gives
\[
 S(W_1W_2)_\tau-H(X_1X_2)_\tau\ge G(A)+G(A').
\]
Adding $H(X_1)+H(X_2)$ proves~\eqref{eq:product-witness}.
\qed

\bibliographystyle{unsrt}
\bibliography{refs}
\end{document}